\documentclass[review]{article}
\usepackage{arxiv}
\usepackage[utf8]{inputenc}
\usepackage{booktabs}

\usepackage{graphicx}
\usepackage{siunitx}
\usepackage{booktabs,caption}
\usepackage[flushleft]{threeparttable}
\usepackage{adjustbox}
\usepackage{siunitx}
\usepackage{amsmath}
\usepackage{lineno}
\usepackage{soul, color} % To highlight texts with the command `hl'
\usepackage[english]{babel}
\usepackage{graphicx}
\usepackage{subcaption}
\usepackage{diagbox}
\usepackage{bbm}
\usepackage{array, booktabs, makecell}
\usepackage{siunitx, mhchem}
\usepackage{xr}

\graphicspath{{Figures/}}

\makeatletter
\def\ps@pprintTitle{%
 \let\@oddhead\@empty
 \let\@evenhead\@empty
 \def\@oddfoot{\centerline{\thepage}}%
 \let\@evenfoot\@oddfoot}

\newcommand*{\addFileDependency}[1]{% argument=file name and extension
  \typeout{(#1)}
  \@addtofilelist{#1}
  \IfFileExists{#1}{}{\typeout{No file #1.}}
}
\makeatother

\newcommand*{\myexternaldocument}[1]{%
    \externaldocument{#1}%
    \addFileDependency{#1.tex}%
    \addFileDependency{#1.aux}%
}

\myexternaldocument{supplementary}
\title{U.S. Power Resilience for 2002--2019}
\author{ Aman Ankit \\
	Department of Industrial and Systems Engineering\\
	University of Washington\\
	Seattle, WA 98105, USA \\
\And 
Zhanlin Liu \\
	Department of Industrial and Systems Engineering\\
	University of Washington\\
	Seattle, WA 98105, USA \\
\And 
Scott B. Miles \\
Department of Human Centered Design \& Engineering \\
	University of Washington\\
	Seattle, WA 98105, USA \\
\And 
Youngjun Choe \\
	Department of Industrial and Systems Engineering\\
	University of Washington\\
	Seattle, WA 98105, USA \\
}
\date{July 2021}
\begin{document}
\maketitle

% \author[*]{Aman Ankit}

% \author[*]{Zhanlin Liu}

% \author[**]{Scott B. Miles}

% \author[*]{Youngjun Choe}
% % \cortext[mycorrespondingauthor]{Corresponding author}
% %  ead{ychoe@uw.edu}

% \affil[*]{Department of Industrial and Systems Engineering, University of Washington, Seattle, WA 98105, USA}
% \affil[**]{Department of Human Centered Design \& Engineering, University of Washington, Seattle, WA 98105, USA}


% \date{August 2020}

% \section{Abstract 150}
\begin{abstract}
Prolonged power outages debilitate the economy and threaten public health. Existing research is generally limited in its scope to a single event, an outage cause, or a region. Here, we provide one of the most comprehensive analyses of U.S. power outages for 2002--2019. We categorized all outage data collected under U.S. federal mandates into four outage causes and computed industry-standard reliability metrics. Our spatiotemporal analysis reveals six of the most resilient U.S. states since 2010, improvement of power resilience against natural hazards in the south and northeast regions, and a disproportionately large number of human attacks for its population in the Western Electricity Coordinating Council region. Our regression analysis identifies several statistically significant predictors and hypotheses for power resilience. Furthermore, we propose a novel framework for analyzing outage data using differential weighting and influential points to better understand power resilience. We share curated data and code as Supplementary Materials.
%(150) Prolonged power outages severely impact thousands of people who depend on electricity for their in-home medical devices. Most research focuses on a single type of power outage such as those caused by natural hazards, leaving room for a comprehensive analysis that studies multiple types. Here, we utilize 16 years of power outage data by categorizing them into four comprehensive categories—natural hazards, human attacks, operational maintenance, and mechanical failures—and analyze it using industry standard IEEE 1366-2012 reliability metrics. We provide clean datasets and code to reproduce or further experiment with the methodologies. Our spatiotemporal analysis reveals six of the most resilient states since 2010 (Alaska, West Virginia, Montana, Wyoming, South Dakota, and Nebraska), that the south and northeast regions have improved their power resilience by X\% following extreme natural weather events, and a negative correlation between the number of unique utility companies in a region and resilience. Our LASSO regression analysis identifies several predictors and hypotheses for future investigation. We further propose a new framework of analyzing power outage data using influential points and weighted data to better understand the resilience of an electrical infrastructure. (190)
\end{abstract}
% \end{frontmatter}
%\linenumbers

\section{Introduction}

Although power outages affect everyone, it severely impacts those whose access to electricity is so critical that a lapse in the electrical infrastructure can be fatal or cause permanent damage \cite{molinari_2017}. For instance, there is an increasing prevalence of diabetes in Puerto Rico  \cite{cdc_2012} and many of them depend on electricity-powered dialysis machines, but when Hurricane Maria struck in 2017, it left virtually all residents without power \cite{mellgard_2019} and was not restored to normal levels until after 10 months  \cite{kwasinski_2019}. Another power outage induced by 2021 winter storms in Texas over two weeks cost over 200 lives \cite{hauser_sandoval_2021}. These incidents solemnly call for a comprehensive analysis of historical U.S. power outage data. This study uses U.S. federal data currently available from 2002 to 2019 to examine the past and identify the trends that remain in force.

%Due to a strong positive correlation between climate change and extreme weather \cite{melilo_2014}, t
There are numerous studies on the impact of extreme weather on above-ground electrical infrastructure  \cite{ward_2013,shafieezadeh_2014,wilbanks_2008,hall_2012,fluke_2017,shandiz2020resilience} or on energy systems in general  \cite{schaeffer2012energy,perera_2020,yalew_2020,tian2021energy, cadini2017modeling, meier2019using} and the aftermath of power outages caused by extreme weather events \cite{abi-samara_2014,nas_2017,ji2016large,dobson2016electricity}. However, these studies mainly focus on outages caused by natural hazards or effects of climate change. In our paper, we expanded this scope to also include power outages caused by human attacks, mechanical failures, and operational maintenance. Studies on cyberattacks and deliberate physical attacks are often qualitative and do not discuss in depth outages caused by natural hazards or mechanical failures \cite{burke_2015,sullivan_2017} or are limited to microgrids \cite{mishra_2020}.

Many studies investigate power outages using mathematical  \cite{ji_2017} and statistical models such as support vector regression  \cite{chen2014short}; support vector machine-random forest  \cite{mukherjee_2018}; ANOVA, ARMA, and Poisson regression  \cite{adderly_2016}; and linear modelling  \cite{larsen_2020}, but they either have a significantly different scope and approach from our work, do not incorporate the Institute of Electrical and Electronics Engineers (IEEE) 1366 reliability metrics, use a smaller data set that does not encompass the entire U.S., contains much less than 18-years of data, or was specific to a certain type of outage. In addition to statistical modeling, understanding the resilience of the country at scale requires a large data set and evaluation of multiple outage causes simultaneously using widely-used metrics in the subject matter.

We conduct an objective evaluation of the U.S. electrical grid in two ways. First, we analyze three established IEEE 1366-2012 reliability metrics---System Average Interruption Duration Index (SAIDI), System Average Interruption Frequency Index (SAIFI), and Customer Average Interruption Duration Index (CAIDI)---for all 50 states over a 18-year time period (2002--2019). This analysis resulted in three key discoveries: 1) we found that, since 2010, Alaska, West Virginia, Montana, Wyoming, South Dakota, and Nebraska are overall the most resilient states for all types of outages, 2) North American Electric Reliability Corporation (NERC) region Western Electricity Coordinating Council (WECC) has a disproportionately large number of human attacks for its population, 3) the south and northeast regions of the U.S. have improved their resilience to extreme weather over time when compared to outages caused by human attacks, mechanical failures, and operational maintenance. Second, our regression analysis of the reliability metrics with 41 explanatory variables revealed two additional discoveries: 4) regions that have fewer unique utility companies are more resilient, 5) numerous strong predictors that will inspire future research hypotheses. We conclude our analysis by proposing a new framework for interpreting the reliability metrics, which includes weighting outages and using influential points to identify anomalies that provide new insight into power resilience.

\section{Methods}\label{sec:methods}
\subsection{Data}
The U.S. Department of Energy (DOE) mandates that all electric and emergency incidents and disturbances be recorded within a certain amount of time (within 1 or 6 hours of incident event depending on the alert criterion established by the DOE   \cite{oe417form}). The DOE collects this information through form OE-417 for national security, energy emergency management responsibilities, and other analytical purposes \cite{oe417data}.
 
OE-417 includes information about the date and time the event began and restored (the time and date of outage and its restoration), the area affected (state and, if applicable, the specific county), the North American Electric Reliability Corporation (NERC) region, alert criteria, event type (cause of outage), the demand loss in megawatts, and the number of customers affected. These values are recorded then uploaded as Excel files to the DOE website  \cite{oe417_annualsummary}. The records date back to 2000, but for this study we are using records from 2002 and beyond because 2000 and 2001 have data in PDF format instead of in Excel or CSV format.
 	
The second source of data used is from the U.S. Energy Information Administration (EIA). The purpose of the EIA is to collect and analyze independent and impartial information regarding energy. The EIA collects surveys on electric power data. A specific form EIA-861 is used to record electric power sales, revenue, and energy efficiency data, customer counts, etc. from each utility company. These files are available in Excel formats from the years 1990--2019  \cite{eia}.
 
OE-417 provides us with the details surrounding the outage including the number of people affected, duration, and cause. EIA-861 provides us with the information regarding how many customers were sold electricity by a breakdown of state. We limited our period from 2002--2019 since 2002 was the oldest published and parseable dataset that the DOE had records for OE-417.
 
The data can be combined to calculate the percentage of customers affected and for how long. Supplementary Tables 2 and 3 show useful information that can be extracted from OE-417 and EIA-861, respectively. Before combining the two datasets using the Area Affected column from OE-417 and the State column from EIA-861, we took several steps to clean the dataset. For the OE-417 dataset, we added a column for time elapsed (days) regarding the duration of the outage and deleted rows where the number of customers affected was missing. After doing this calculation, we observed that some time elapsed values were negative. We reviewed these observations on a case-by-case basis. If we determined with high confidence that it was possible for the operator to mistakenly input AM when they meant PM or vice versa we manually made the proper adjustment to reflect the possible actual value otherwise the observation was removed. We verified these with reported news articles or alerts around the time of the event. There were some observations in which the number of customers affected were negative. We found no explanation for negative values in this column, so we made the assumption that such observations were meant to be positive but were mistakenly inputted as negative.
 
For OE-417 2002 through 2019, the Event Type column had 46 unique causes. These causes were simplified to four overarching reasons that we determined were still able to capture the category of the cause. The first category was natural hazards, which included all outages directly caused by any natural event such as high winds, damage due to trees, and any type of severe weather and outages indirectly caused by natural events such as voltage reduction and fuel supply emergencies due to unfavorable weather. The second category was mechanical failure, which included all outages caused by fuel supply emergencies due to limited resources, distribution interruptions, transmission interruption, generation inadequacy, and units or transmission trips. The third category was human attack, which included outages caused by vandalism, cyber-attacks, and sabotage. The fourth category was operational maintenance caused by deliberate outages due to islanding, load shedding, public appeal to reduce electricity usage, planned outages, or operational failures caused by uncontrollable outages due to a lack of preparedness in operations.

\subsection{Spatiotemporal Analysis}
Once we combined the OE-417 dataset with the EIA-861 dataset, we were able to calculate the percentage of customers affected. To explore the reliability based on different outage causes, we calculated the System Average Interruption Duration Index (SAIDI), System Average Interruption Frequency Index (SAIFI), and Customer Average Interruption Duration Index (CAIDI) for each state considering all causes of outage and for each state considering each cause of outage. IEEE Standard 1366 establishes a list of reliability metrics (including the three most popular metrics SAIDI, SAIFI, and CAIDI) to provide a standard way for all electric utilities to measure electrical reliability and ensure consistency and compatibility. The existing literature frequently uses the three metrics, but our approach is unique in that we analyze reliability based on four comprehensive causes. This helps give new insight into the electrical grid as it allows for the analysis at a clustered level. It is also important to mention that as of January 2017 “only 33 percent of utilities report [SAIDI and SAIFI] statistics”  \cite{qer_017}which prompted us to calculate these metrics for all states as well. The reliability metrics are calculated each year, but we also calculated them for longer time periods as well for spatiotemporal analysis and trend identification. With our 18 years of data from 2002 – 2019, we separated the data into two parts of 9 years each: 2002 – 2011 (year range 1) and 2011 – 2019 (year range 2).

\subsection{Regression Analysis}
We used a total of 41 variables as the potential predictors in our least absolute shrinkage and selection operator (LASSO) regression model \cite{tibshirani1996regression} to select the variables that best predict the three metrics (SAIDI, SAIFI, and CAIDI) based on the 10-fold cross-validation. The data collected by Mukherjee et al. contains 55 variables that pertain to major power outage events \cite{data_2018}. We used 39 of the variables provided by Mukherjee et al. and combined them with an additional two predictors: number of unique companies (from EIA-861) and land area in square miles (from the U.S. Census Bureau). LASSO was chosen over other variable selection methods due to its interpretability and general robustness against collinearity between predictors. We calculated the three metrics for each state and each outage cause category. We then used the selected variables as the predictors in an ordinary least-squares (OLS) regression model and selected the variables that had $p$-values less than 0.10 before refitting the final OLS model. The chosen $p$-value threshold of 0.10 allows us to identify potentially meaningful predictors of the reliability metrics. The regression coefficients, intercepts, standard errors, $t$-values, and $p$-values of each model are summarized in Supplementary Tables {4--6}.

\section{Spatiotemporal Analysis of U.S. Power Resilience}
This section analyzes how power outage patterns vary across states over time. To organize all power outages into a comprehensive analysis, we simplified 46 reasons for power outages into four overarching causes: natural (including but not limited to hazards and weather events), mechanical failure, human attack, and operational maintenance (see Section~\ref{sec:methods} for definitions). To explore the resilience based on the four different causes we calculated three IEEE Standard 1366 reliability metrics (SAIDI, SAIFI, and CAIDI) for each state considering all causes of outage and for each state considering each cause of outage.
%The standard reliability metrics are often referenced in existing literature, but our analysis is unique in that it provides new insight into the electrical grid at an aggregate level (by state and cause).

\begin{figure}[htp]
    \centering
    \includegraphics[width=0.95\textwidth]{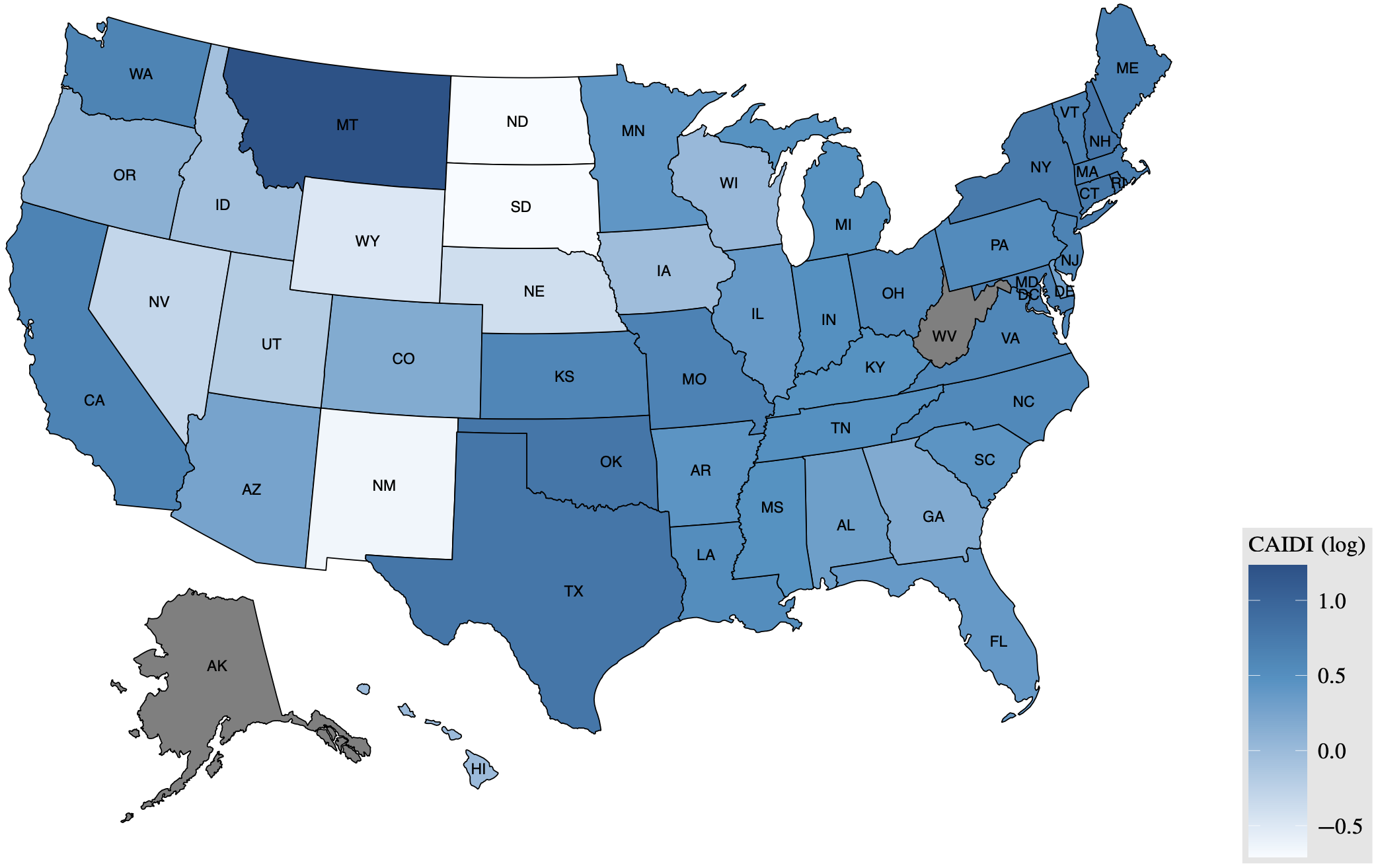}
    \caption{CAIDI values for each state for power outages caused by natural hazards on a log scale (2002--2019).}
    \label{fig:1}
\end{figure}

Figure \ref{fig:1} is a map of the CAIDI values on a log scale for each state for the natural hazard type. Since CAIDI measures the average duration length per outage, a lower CAIDI value translates to a more resilient infrastructure. Immediately, we notice there are two gray states: West Virginia and Alaska. Although the gray states are not on the reliability metric scale (i.e. not a sufficient number of outage observations in OE-417 to calculate a CAIDI value), note that the lack of a CAIDI value itself is a measurement of reliability. If there were no outages recorded then the electrical structure was robust enough to withstand disruptive natural hazards such that outages never occurred or that the outages were insignificant (e.g., small flicker in power) because the infrastructure was strong enough to not be affected and was thus not reported in OE-417 (see Supplementary Note 1.1 for more details including federal reporting mandates on OE-417).

% FIGURE 2A CAPTION: CAIDI values for each state all types of power outages in 2002 -- 2009 on a log scale.
% FIGURE 2B CAPTION: CAIDI values for each state all types of power outages in 2010 -- 2017 on a log scale.
% insert subcaption (a) and (b)
\begin{figure}[htp]
    \centering
    \includegraphics[width=15cm]{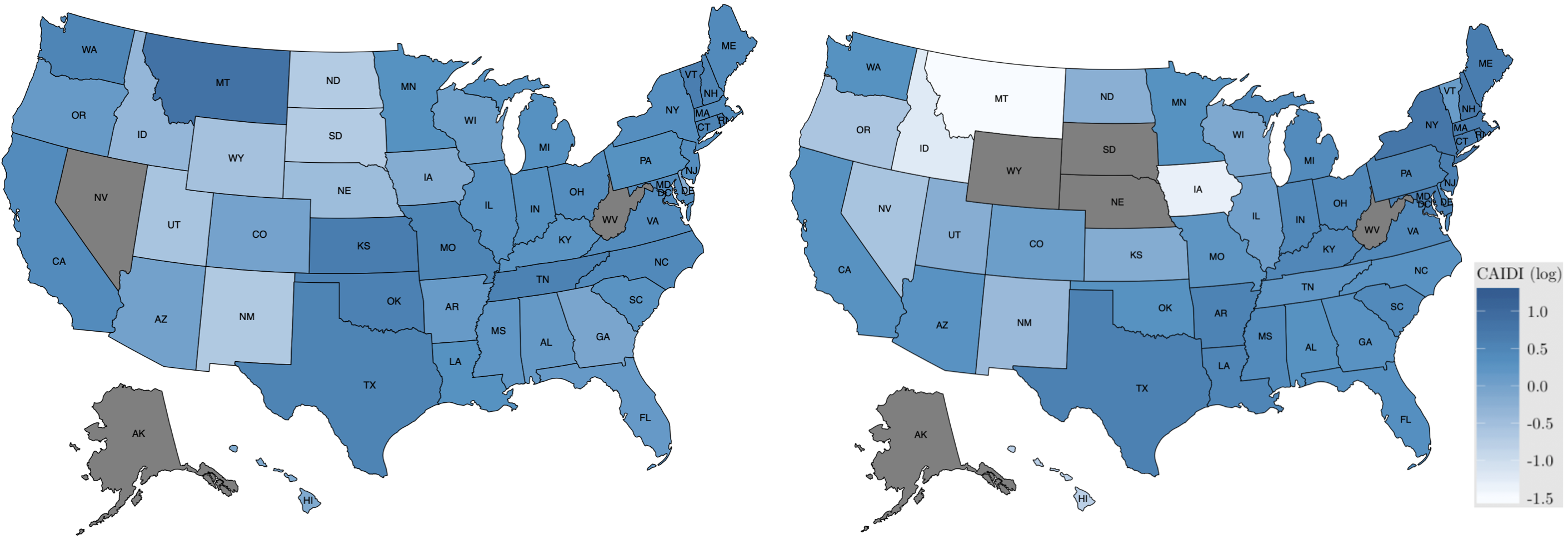}
    \caption{CAIDI values for each state considering all types of power outages in 2002--2010 (a) and 2011--2019 (b) on a log scale.}
    \label{fig:2}
\end{figure}

Figure \ref{fig:2} shows CAIDI values for each state over two large time periods. 
%We can extend the previous idea regarding one possible interpretation of gray states and apply it here as well. We see that 
Alaska and West Virginia remained gray indicating that through the 18 years, there were no outages severe enough to warrant reporting the outage at a federal level. Nevada in Figure \ref{fig:2}b is now light blue from gray in Figure \ref{fig:2}a indicating a decline in reliability. This change was a result of 13 human attack incidents, two natural hazards, and two incidents for operational maintenance. Nevada is a part of the Western Electricity Coordinating Council (WECC) which sees the highest number of human attacks (38.1\% of the human attacks in all NERC regions occurred in WECC) as shown in Table \ref{Table 1}.

\begin{table}[ht]
\centering
\caption{Number of power outages caused by human attacks per NERC region}
\label{Table 1}
\begin{tabular}{cc}
\toprule
\textbf{NERC Region} & \textbf{\thead{Number of Outages\\ Caused by Human Attack}} \\ \midrule
WECC                 & 80                                                \\
RFC                  & 64                                                \\
NPCC                 & 36                                                \\
SERC                 & 11                                                \\
TRE                  & 7                                                 \\
MRO                  & 4                                                 \\
SPP                  & 4                                                 \\
FRCC                 & 2                                                 \\
MISO                 & 2                                                 \\
HI                   & 0                                                 \\
AK                   & 0                                                 \\ \bottomrule
\end{tabular}
\end{table}

One may interpret this as a consequence of the large area that WECC oversees (Supplementary Figure 1) and suspect that a higher population will naturally result in more frequent outages to be caused by human attacks. However, this is not true when comparing the population sizes: from 2011 – 2019, WECC oversaw 32,820,380 electrical customers per year on average and SERC oversaw 32,852,013 customers per year on average and yet only had 11 human attacks (5.2\% of the total human attacks in all U.S. NERC regions).

\begin{table}[ht]
\centering
\caption{Number of outages by NERC region per type of cause}
\label{Table 2}
\begin{tabular}{c c c}
\toprule
\textbf{Cause} & \textbf{SERC} & \textbf{WECC} \\ \hline
Human attack               & 11            & 80            \\
Mechanical failure         & 11             & 17            \\
Natural hazard                    & 495           & 138           \\
Operations                 & 29            & 89            \\ \hline
\textbf{\textit{Total}}                      & \textbf{\textit{546}}           & \textbf{\textit{324}}
   \\\hline
\end{tabular}
\end{table}

In addition, Table \ref{Table 2} shows that SERC has 62.8\% more outages of all kinds than WECC (546 compared to 324). Out of all outages that occur, 24.69\% of outages were due to human attacks in WECC, but only 2.02\% of outages were due to human attacks in SERC. It is clear that WECC has a disproportionately large number of outages caused by human attacks. 
%leading us to believe that there must be something unique to the WECC region that correlates with a higher probability of human attacks that resulted in decreased resilience in Nevada. 
One possible explanation for this may be that WECC does not prioritize placing preventative measures for physical and cyber-attacks as much as SERC. The reasoning for this can be two-fold. (Hypothesis 1) By deprioritizing the allocation of initial preventative measures that would have stopped outages occurring from human attacks, they redirect the resources to prevent other types of causes. A reason for this could be that outages caused by human attacks are typically restored more rapidly than outages caused by other types. (Hypothesis 2) WECC conducted its own analysis and determined that it is increasingly difficult to prevent human attacks and thus resorted to allocating resources that would shorten the restoration time instead of allocating resources and money for initial preventive measures such as undergrounding since it is not always cost effective \cite{larsen_cost_2018}. Both hypotheses could be investigated using financial information and more detailed reports of the systems and rules in place, but Table \ref{Table 3} shows that human attacks in WECC indeed recover faster than any other type of attack as indicated by the CAIDI value. Easily accessible and well-curated data would aid in determining exact root causes instead of hypothesizing. A lack of data makes it difficult to come to exact conclusions and limits further analyses \cite{rocchetta_2018}.

\begin{table}[]
\centering
\caption{CAIDI values per outage type for WECC 2002--2019}
\label{Table 3}
\begin{tabular}{cc}
\toprule
\textbf{Reason for Outage} & \textbf{CAIDI} \\ \midrule
Human attack       & 0.0728 \\
Mechanical failure & 0.3296 \\
Natural hazard     & 4.0848 \\
Operations         & 1.2837 \\ \bottomrule
\end{tabular}
\end{table}

\begin{figure}[ht]
    \centering
    \includegraphics[width=0.95\textwidth]{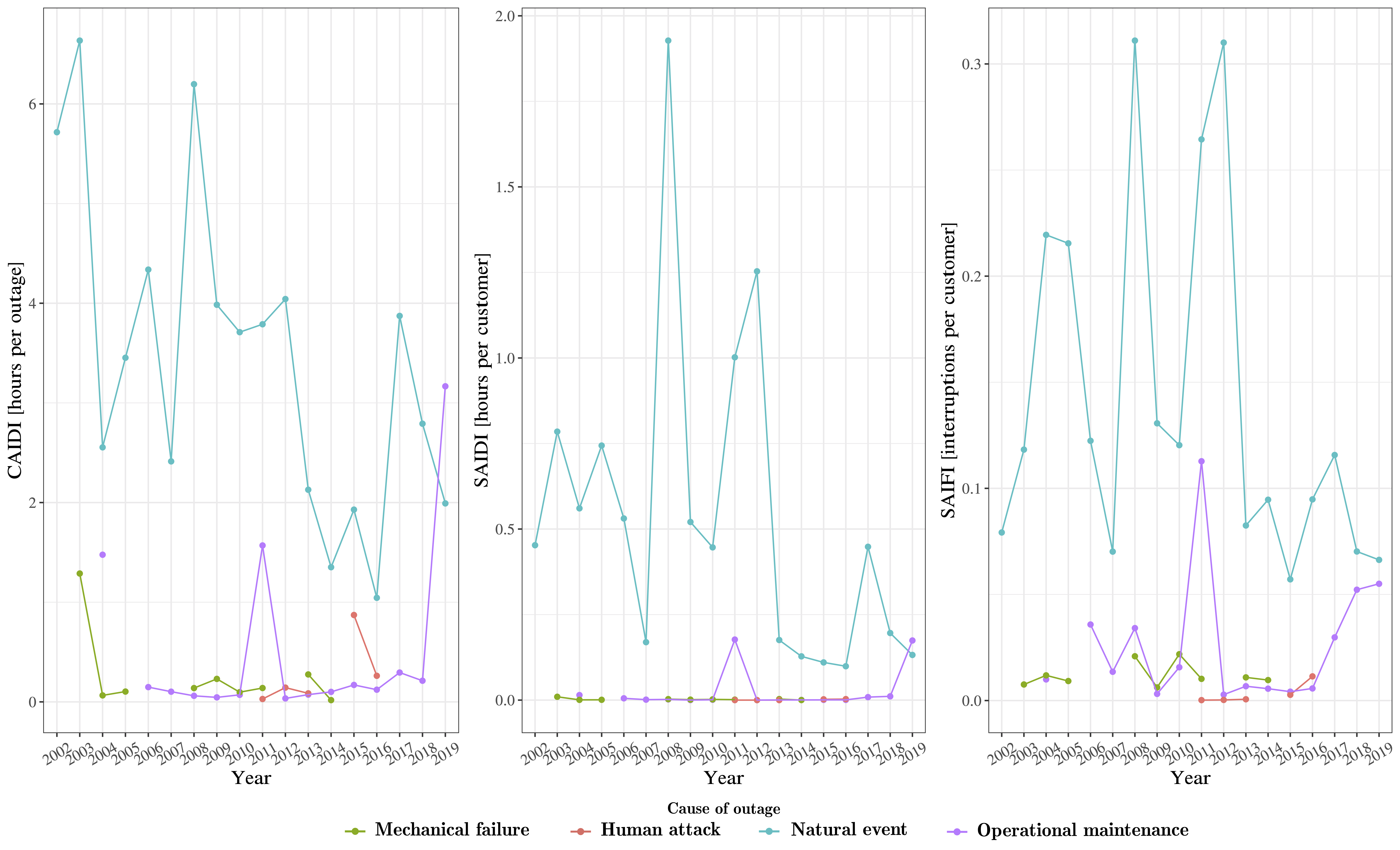}
    \caption{CAIDI (a), SAIDI (b), and SAIFI (c) for 2002--2019 by outage type.} %(b) SAIDI values for 2002 — 2017 by outage type (c) SAIFI values for 2002 — 2017 by outage type}
    \label{fig:3a3b3c}
\end{figure}

In Figure \ref{fig:3a3b3c}a, we observe a sudden increase of outages by natural hazards in 2017, which is an outlier from the decreasing trend in the previous eight years. This outlier was caused by Hurricanes Harvey, Irma, Lee, and Maria. The last major increase most similar to that of 2017 was in 2008 due to the 2008 Super Tuesday tornado outbreak, Hurricane Ike, and ice storms. The events in 2008 and 2017 occurred in the same south and northeast regions of the U.S. \cite{us_regions} A difference between 2008 and 2017 is that in 2008 SAIDI and SAIFI (Figures \ref{fig:3a3b3c}b and \ref{fig:3a3b3c}c) suddenly increase by 1038\% and 343\%, respectively, but in 2017 SAIDI and SAIFI increase by 353\% and 22\%, respectively, while in 2008 CAIDI increased by 157\% and 271\% in 2017. Although the increase in CAIDI is higher in 2017 than it was in 2008, the lower percentage increase in SAIDI and SAIFI in 2017 suggests an improvement in the infrastructure and/or preparedness in those particular regions. 
%that attribute to lower SAIDI and SAIFI values during natural hazard-induced outages.

We hypothesize that this improvement is attributed to 
%many things, but perhaps the most prominent evidence that explains this improvement is 
the decline in the number of unique electrical companies.
%. In our regression analysis, we found that the number of unique companies was 
%, so a decline in the number of companies over time would indicate an improvement in CAIDI. Observing a decrease in the value of CAIDI further supports the claim made in our regression analysis discussion as 
 The number of unique companies declined from 3421 in 2002 to 3259 in 2017 (a 4.73\% decrease) 
 %In 2002 there were 3421 companies and in 2017 there were 3259 companies (a 4.73\% decrease). 
 despite the electrical customer base growing from 138,434,951 customers in 2002 to 164,475,172 customers in 2017 (an 18.81\% increase) in the U.S. 
 %, a decline in the number of unique companies substantiates our hypothesis that a large number of unique companies indicates a large number of smaller utility companies with presumably smaller budgets. 
 When small companies are consolidated with larger ones, it enables economies of scale (e.g., greater availability of specialized repair crew and better optimization of resource allocation). 
 %allows for higher resource availability, expertise, and tooling equipment that was previously inaccessible.

\section{Regression Analysis of U.S. Power Resilience}\label{sec:regression}

We conducted a regression analysis to identify multiple hypotheses on U.S. power resilience based on 41 potential predictors. Recall that a regression analysis can identify a significant \textit{association} between a predictor and a response variable (while controlling for other predictors’ associations with the response variable). The significant associations reported below should not be construed as definite \textit{causation}. Our stated hypotheses implying causal relationships are meant to inspire future work for testing the hypotheses or investigating the associations further (e.g., collinearity with other predictors not included in our analysis). In this study, the response variables are IEEE reliability metrics for each state
\begin{itemize}
    \item considering all causes of outage (denoted, for example, as ``SAIFI state") or
    \item considering each cause of outage (denoted, for example, as ``SAIDI mechanical" for SAIDI calculated with respect to mechanical failures only).
\end{itemize}% or for each metric and cause combination . 
Below we discuss major findings and defer additional analysis results to {Supplementary Notes 1.2 - 1.4}.

Our analysis found that SAIDI mechanical was positively correlated with the number of residential customers (regression coefficient: \num{9.739e-10}, standard error: \num{4.208e-10}, $p$-value: 0.02510).  A 2013 White House report found that “the response time of grid operators to mechanical failures is constrained by a lack of automated sensors” \cite{wh_energy_2013}. Aged automated sensors or absence of automated sensors can result in slower or even missed detections. One hypothesis that relates lack of functioning automated sensors with increasing residential customers is that prioritization of new technologies might get delayed in order to ensure that it can properly support a large residential customer base and properly address customer complaints or due to public hesitation despite various benefits \cite{wang_2018}. For example, a public utility company in Seattle, WA (population in 2019: 744,955 \cite{bureau_2019}) is deploying and installing smart meters in all houses in Seattle. One new feature in the smart meter or “Advanced Metering” system allows for power outage detection \cite{scl}. The implementation of this new technology is delayed partially due to a large increase in the residential population resulting in an underestimated initial budget and thus delaying the full rollout to stay under budget \cite{kroman_2017}. Furthermore, automated sensors using outdated hardware designs may not have the proper redundancies of key elements built in to withstand some of the mechanical failures and prevent blackouts  \cite{mazur_2019}.

For SAIFI state, states with higher urban population and density have fewer interruptions per customer. The two significant predictors chosen by LASSO were population percentage in urban areas (regression coefficient: \num{-3.2903e-2}, standard error: \num{1.4777e-2}, $p$-value: 0.03091) and population density of urban clusters (regression coefficient: \num{-2.2244e-3}, standard error: \num{7.104e-4}, $p$-value: 0.00302).  Both predictors had a negative correlation with response variable.
One hypothesis for this relationship can be attributed to the fact that states that have higher urban population will be states where utility companies are more likely to have their electrical infrastructure underground making it less susceptible to all kinds of outages. It is widely known that undergrounding is more expensive than overhead power lines, so utility companies will look to place power lines underground only where there is a high chance of recouping the costs via taxes or electrical bills. There is strong evidence that links the growth of urbanized regions with an increase in economic activity, so it is more likely that utility companies place their power lines underground in urban areas \cite{inklaar_2018}. Highly urbanized areas such as Seattle also offer undergrounding services demonstrating that as the economic activity grows, utility companies may see increased demand for undergrounding \cite{scl_undergrounding}. Restoration resources would also be abundant in places with higher urban populations because of the larger impact outages have due to higher population density allowing for faster recovery than states with lower ``operational agility"  \cite{jufri_2019}. Recall that the equation for calculating SAIFI had the total number of customers served in the denominator, which means that in regions that have large population (large total number of customers served), such as urban areas, will result in lower SAIFI values. Because of large populations in urban areas, we also expect their electrical infrastructures to be large enough or larger to support the growing population and demand. Due to their more advanced infrastructure, urban areas may use their power transmission systems to sell their power to supplement utility companies in nearby regions \cite{cadini2017modeling}. This secondary income would further incentivize investing in robust restoration resources because it may also affect nearby regions that rely on buying electricity.

SAIFI mechanical had the percentage of inland water area as its most significant predictor (regression coefficient: \num{2.703e-3}, standard error: \num{1.033e-3}, $p$-value: \num{0.0119}). This relationship may be attributed to corrosion and debris flow that can cause the electrical infrastructure to require replacement or cleaning. For example, floods can affect the electrical equipment, and they may still be functional at the time of the flood, but the damage could be realized weeks or months later due to corrosion thus classifying it as a mechanical failure versus a natural hazard-induced failure. Corroded material may also have a cascading effect \cite{graham_2010} which can affect a large network of customers, hence increasing the SAIFI values.

CAIDI natural's most significant predictor was 
the population density of rural areas (regression coefficient: \num{0.02044}, standard error: \num{0.01186}, $p$-value: \num{0.09140}).
The positive association between the population density of rural areas and CAIDI natural may be attributed to overhead distribution networks, which are common in rural areas and vulnerable to wind hazards, especially, in close proximity to trees \cite{li_2010}. Furthermore, if the rural population density was low then it would not be a significant enough of an outage to make an impact to the CAIDI metric and thus would not be present as a significant predictor in LASSO. Since rural areas are more likely to have radial distribution systems \cite{mukherjee_2018}, which are more prone to failures, an increasing rural population density would increase the CAIDI metric. 

\section{A Novel Power Resilience Analysis Framework}\label{sec:framework}

This section discusses a novel framework to evaluate power resilience. All metrics in IEEE Standard 1366 are calculated using unweighted data without regard for differences between regions, socio-economic factors, regulatory standards, system configuration, customer density, hazard exposure, and various other factors that differentially affect the metrics \cite{qer_017}. For example, it is not fair to measure power resilience in all regions equally if the damage and economic loss during the outage affects different regions disproportionately \cite{schmidthaler_2016}. We propose a flexible framework to assign any type of quantitative weight to each observation or groups of observations (all observations for a state) to account for the differences and standardize the reliability metrics. This would also allow metrics to be divided into categories such as ``consequence-based" or ``service-based" \cite{shandiz2020resilience}. The objective of this framework is to also address a shortcoming of the 2.5 Beta Method used to identify Major Event Days (MEDs), which serve as identifiers for unexpected large-scale events that have a low probability of occurring based on past data \cite{1366}. One shortcoming in the 2.5 Beta Method is the degree of influence outliers can have on the value of the threshold resulting in events that were major in one year, but considered normal the next \cite{hann_2011} or vice versa. Our proposal of using influential points in combination with weighting the outage data mitigates this issue.

\begin{figure}[ht]
    \centering
    \includegraphics[width=0.95\textwidth]{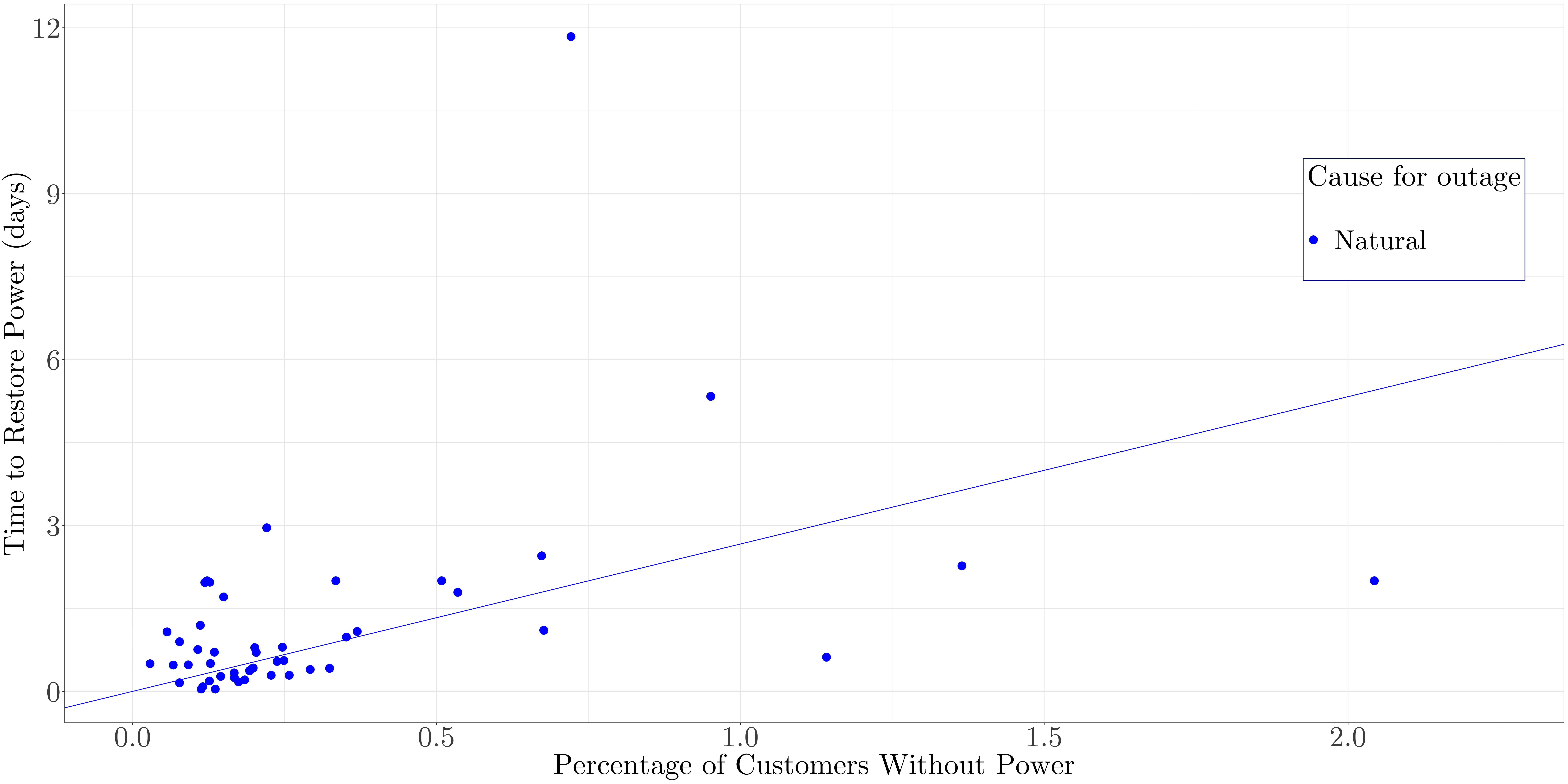}
    \caption{Natural hazard-induced power outage events with a regression line for the state of Alabama for 2002--2019.}
    \label{tab:reliability via regression}
\end{figure}

This framework is grounded in an intuitive visualization of outage data (publicly available according to U.S. regulations).
This visualization plots each observation on a graph as demonstrated in Figure \ref{tab:reliability via regression} where the duration of the outage (time elapsed in days) is on the $y$-axis and the percentage of customers without power is on the $x$-axis. Then, the slope of the regression line created by ordinary least squares is related to CAIDI $(=\sum r_i N_i / \sum N_i)$, SAIFI $(=\sum N_i / N_T)$, and SAIDI $(=\sum r_i N_i / N_T)$as follows:

\begin{linenomath}
\begin{align*}
    \textrm{Slope} &=  \frac{\sum_{i=1}^{n} r_i  N_i}{\sum_{i=1}^{n}N_i^2} N_T\\
    &=  \left(\frac{\sum_{i=1}^{n} r_i  N_i}{\sum_{i=1}^{n}N_i} \right) \left(\frac{\sum_{i=1}^{n}  N_i}{N_T} \right) \frac{N_T}{\sum_{i=1}^{n}N_i^2} N_T\\
    &= \left(\textrm{CAIDI} \right) \left( \textrm{SAIFI}\right)  \frac{1}{\sum_{i=1}^{n}\left(N_i/N_T\right)^2} \\
    &= \left(\textrm{SAIDI} \right)  \frac{1}{\sum_{i=1}^{n}\left(N_i/N_T\right)^2},
\end{align*}
\end{linenomath}

where $n$ is the total number of events, $r_i$ is the restoration time of the $i$-th event in days, $N_i$ is the total number of customers interrupted by the $i$-th event, and $N_T$ is the total number of customers served in the studied region.
Thus, the regression slope characterizes the electrical reliability. Note that $N_i/N_T$ is the proportion of customers interrupted by the $i$-th event in the studied region. This normalized quantity facilitates comparisons across differently populated regions. Its larger value influences the slope more. Similarly, the larger the $r_i$, the greater its influence to SAIDI and thus the slope. 
%Note the following relation between CAIDI, SAIDI, and SAIFI ([EQUATION]) makes the slope related to all three metrics. 

 When examining the data in this manner, we can apply a different weight on the $i$-th observation for $i=1,2,\ldots,n$ to reflect its normalized effect on the slope and thus the reliability according to various weighting schemes (e.g., the weight may be proportional to the proportion of population older than 65 in the affected region to reflect social vulnerability).

% \begin{table}[]
% \centering
% \begin{threeparttable}
% \caption{Measure of influence using various methods for California for 2002--2017}
% \label{tab:influence}
% \begin{tabular}{SSSSSS[table-format=3.2]}
% \toprule
% \textbf{Observation} & \textbf{\thead{Difference\\of Betas}} & \textbf{\thead{Difference\\of Fit}} & \textbf{\thead{Coefficient of\\Variation Ratio}} & \textbf{Cook's Distance} & \textbf{\thead{Hat Diagonal\\Values}} \\ \midrule
% 1  & -0.02 & -0.02   & 1.06 * & 0.00   & 0.04   \\
% 11 & -0.31 & -0.31   & 1.17 * & 0.10   & 0.14 * \\
% 14 & -0.01 & -0.01   & 1.10 * & 0.00   & 0.08 * \\
% 21 & -0.02 & -0.02   & 1.53 * & 0.00   & 0.34 * \\
% 29 & 0.58  & 0.58 *  & 1.15 * & 0.33   & 0.14 * \\
% 41 & 0.06  & 0.06    & 0.88 * & 0.00   & 0.00   \\
% 47 & -0.83 & -0.83 * & 1.02   & 0.64 * & 0.10 * \\
% 54 & 0.05  & 0.05    & 0.94 * & 0.00   & 0.00   \\
% 59 & 0.08  & 0.08    & 0.94 * & 0.01   & 0.00   \\
% 62 & 0.10  & 0.10    & 0.81 * & 0.01   & 0.00   \\ \bottomrule
% \end{tabular}
%   \begin{tablenotes}
%       \small
%       \item \textit{Note}: For each measure of influence, observations with a starred number were deemed influential.
%     \end{tablenotes}
%   \end{threeparttable}
% \end{table}

%A prominent example of this could be in California. 
Using this framework conducive to regression modeling, we can algorithmically evaluate the influence of each point (or event) on the regression slope (or electrical reliability). To illustrate this, we used California data. As shown in Supplementary Table 1, we determined influential points using commonly used measures in the regression literature. %(implemented in the `influence.measures' function in the R Stats package).
For example, with Cook’s Distance (which is arguably the most widely used measure in practice), there is one influential point in the dataset of 62 observations (see Supplementary Figure 2). This observation corresponds to an event that occurred on May 16, 2014 in San Diego and Orange Counties. The outage due to a natural hazard (wildfire) affected 1,400,000 people and lasted approximately 0.4 hours. Table \ref{Table 8} shows the calculated reliability metrics with and without the identified influential point where the total number of outage events was three in 2014.

\begin{table}[]
\centering
\caption{Three reliability metrics with and without the influential point in California where three outage events due to natural hazard were observed in 2014.}
\label{Table 8}
\begin{adjustbox}{width=0.98\textwidth}
\small
\begin{tabular}{cccc}
\toprule
\textbf{IEEE 1366 reliability metric} & \textbf{\thead{With \\influential point}} & \textbf{\thead{Without \\influential point}} & \textbf{\% change} \\ \midrule
SAIDI {[}hours per customer{]}         & 0.0559 & 0.0171 & \textbf{- 69.41\%} \\
SAIFI {[}interruptions per customer{]} & 0.1009 & 0.0100 & \textbf{- 90.09\%} \\
CAIDI {[}hours per outage{]}           & 0.5551 & 1.702  & \textbf{+ 206.6\%} \\ \bottomrule
\end{tabular}
\end{adjustbox}
\end{table}

We might expect that upon removal of an influential point (i.e., a significant power outage event that was deemed the worst in its group), the CAIDI would decrease. If the MED algorithm were not able to identify this observation as an outlier due to an abnormally high threshold value caused by outliers in the previous year, then we would not know the impact of the event in the reliability calculation as the influential point had the shortest outage duration, but the highest number of customers affected of all observations in California in 2014 caused by natural hazards. Influential points serve as a robust method for identifying high impact events that deviate from the norm while retaining the purpose of using past observations to measure a typical baseline as used in the MED algorithm.

Examining power outage using weighted data is advantageous for a number of reasons. For example, socio-economic factors of different regions can normalize observations. Normalizing can help give further insight into different regions' electrical infrastructures because some observations may have been previously considered noisy or outliers but in reality were a true reflection of the electrical system. Another possible factor to weight the observations by can be a state or NERC region-specific metric like per capita real Gross State Product (GSP), which would normalize economic differences between the regions.

\section{Conclusion}

The hypotheses stated in Section~\ref{sec:regression} in combination with the proposed framework for measuring reliability in Section~\ref{sec:framework} established fertile grounds for future research. Our spatiotemporal and regression analyses revealed power resilience trends and their potential determinants. These analyses also raised questions that may be addressed by future studies. Future work can adopt and expand on our methods to better understand the in-depth effects of population size and density, GSP, number of utility companies in a region, budget allocation, outage restoration procedures, and unique features of a region. Once our hypotheses are tested, policymakers, regulatory and governmental agencies, and utility companies will have a better understanding of creating policies and procedures that result in fewer outages for smaller periods of time. Furthermore, while this study focused on the U.S., similar data in other countries can allow researchers and policymakers to compare infrastructures across countries, creating the opportunity to learn from one another. Our study further confirmed how a lack of standard in data collection can be a hindrance in understanding power resilience, and we hope that decision-makers will consider our recommendations to improve the quality of data collection and the power resilience.

%We hope that our analysis of power outage data lays a foundation for future researchers by demonstrating one way to include large amounts of power outage data points while still conducting a comprehensive analysis. 

\section*{Data Availability} %All manuscripts reporting original research published in Nature journals must include a data availability statement
%https://www.nature.com/nature-research/editorial-policies/reporting-standards 
All the curated data and R code of this study are shared as Supplementary Materials to ensure full reproducibility.

\section*{Acknowledgement}
This work was supported by the National Science Foundation (NSF grant CMMI-1824681).

\bibliographystyle{unsrt}
\bibliography{references}

\begin{thebibliography}{10}

\bibitem{molinari_2017}
N.~A.~M. Molinari, B.~Chen, N.~Krishna, and T.~Morris.
\newblock Who's at risk when the power goes out? {The} at-home
  electricity-dependent population in the {United States}, 2012.
\newblock {\em Journal of Public Health Management and Practice},
  23(2):152--159, 2017.

\bibitem{cdc_2012}
Centers for Disease~Control and Prevention (CDC).
\newblock Increasing prevalence of diagnosed diabetes--united states and puerto
  rico, 1995-2010.
\newblock {\em MMWR. Morbidity and mortality weekly report}, 61(45):918, 2012.

\bibitem{mellgard_2019}
G.~Mellgard, D.~Abramson, C.~Okamura, and H.~Weerahandi.
\newblock Hurricanes and healthcare: a case report on the influences of
  hurricane maria and managed medicare in treating a puerto rican resident.
\newblock {\em {BMC} Health Services Research}, 19(1), November 2019.

\bibitem{kwasinski_2019}
A.~{Kwasinski}, F.~{Andrade}, M.~J. {Castro-Sitiriche}, and
  E.~{O’Neill-Carrillo}.
\newblock Hurricane maria effects on puerto rico electric power infrastructure.
\newblock {\em IEEE Power and Energy Technology Systems Journal}, 6(1):85--94,
  March 2019.

\bibitem{hauser_sandoval_2021}
Christine Hauser and Edgar Sandoval.
\newblock Death toll from texas winter storm continues to rise.
\newblock {\em The New York Times}, Jul 2021.
\newblock Accessed: 2021-7-15.

\bibitem{ward_2013}
D.~M. Ward.
\newblock The effect of weather on grid systems and the reliability of
  electricity supply.
\newblock {\em Climatic Change}, 121:103--113, September 2013.

\bibitem{shafieezadeh_2014}
A.~{Shafieezadeh}, U.~P. {Onyewuchi}, M.~M. {Begovic}, and R.~{DesRoches}.
\newblock Age-dependent fragility models of utility wood poles in power
  distribution networks against extreme wind hazards.
\newblock {\em IEEE Transactions on Power Delivery}, 29:131--139, Feb 2014.

\bibitem{wilbanks_2008}
T.~Wilbanks, V.~Bhatt, D.~Bilello, S.~Bull, and J.~Ekmann.
\newblock Effects of climate change on energy production and use in the {United
  States}.
\newblock {\em US Department of Energy Publications}, 12, 2008.

\bibitem{hall_2012}
K.~L. Hall.
\newblock Out of sight, out of mind: An updated study on the undergrounding of
  overhead power lines.
\newblock Technical report, Edison Electric Institute, 2017.

\bibitem{fluke_2017}
C.~{Fluke}, R.~{Walton}, and S.~{De Merritt}.
\newblock System modernization and reliability: A transition to underground.
\newblock In {\em 2017 IEEE Rural Electric Power Conference (REPC)}, pages
  61--65, April 2017.

\bibitem{shandiz2020resilience}
Saeid~Charani Shandiz, Greg Foliente, Behzad Rismanchi, Amanda Wachtel, and
  Robert~F Jeffers.
\newblock Resilience framework and metrics for energy master planning of
  communities.
\newblock {\em Energy}, 203:117856, 2020.

\bibitem{schaeffer2012energy}
Roberto Schaeffer, Alexandre~Salem Szklo, Andr{\'e} Frossard~Pereira de~Lucena,
  Bruno Soares Moreira~Cesar Borba, Larissa Pinheiro~Pupo Nogueira,
  Fernanda~Pereira Fleming, Alberto Troccoli, Mike Harrison, and
  Mohammed~Sadeck Boulahya.
\newblock Energy sector vulnerability to climate change: A review.
\newblock {\em Energy}, 38(1):1--12, 2012.

\bibitem{perera_2020}
A.~T.~D. Perera, V.~M. Nik, D.~Chen, J.~L. Scartezzini, and T.~Hong.
\newblock Quantifying the impacts of climate change and extreme climate events
  on energy systems.
\newblock {\em Nature Energy}, 5(2):150--159, February 2020.

\bibitem{yalew_2020}
S.~G. Yalew, M.~T.~H. van Vliet, D.~E. H.~J. Gernaat, F.~Ludwig, A.~Miara,
  et~al.
\newblock Impacts of climate change on energy systems in global and regional
  scenarios.
\newblock {\em Nature Energy}, 5(10):794--802, August 2020.

\bibitem{tian2021energy}
Man-Wen Tian and Pouyan Talebizadehsardari.
\newblock Energy cost and efficiency analysis of building resilience against
  power outage by shared parking station for electric vehicles and demand
  response program.
\newblock {\em Energy}, 215:119058, 2021.

\bibitem{cadini2017modeling}
Francesco Cadini, Gian~Luca Agliardi, and Enrico Zio.
\newblock A modeling and simulation framework for the reliability/availability
  assessment of a power transmission grid subject to cascading failures under
  extreme weather conditions.
\newblock {\em Applied Energy}, 185:267--279, 2017.

\bibitem{meier2019using}
Alan Meier, Tsuyoshi Ueno, and Marco Pritoni.
\newblock Using data from connected thermostats to track large power outages in
  the united states.
\newblock {\em Applied Energy}, 256:113940, 2019.

\bibitem{abi-samara_2014}
N.~{Abi-Samra}, J.~{McConnach}, S.~{Mukhopadhyay}, and B.~{Wojszczyk}.
\newblock When the bough breaks: Managing extreme weather events affecting
  electrical power grids.
\newblock {\em IEEE Power and Energy Magazine}, 12(5):61--65, Sep. 2014.

\bibitem{nas_2017}
{National Academies of Sciences, Engineering, and Medicine}.
\newblock {\em Enhancing the resilience of the nation's electricity system}.
\newblock National Academies Press, 2017.

\bibitem{ji2016large}
C.~Ji, Y.~Wei, H.~Mei, J.~Calzada, M.~Carey, et~al.
\newblock Large-scale data analysis of power grid resilience across multiple
  {US} service regions.
\newblock {\em Nature Energy}, 1(5), April 2016.

\bibitem{dobson2016electricity}
I.~Dobson.
\newblock Electricity grid: When the lights go out.
\newblock {\em Nature Energy}, 1(5):1--2, 2016.

\bibitem{burke_2015}
S.~{Burke} and E.~{Schneider}.
\newblock Enemy number one for the electric grid: Mother nature.
\newblock {\em {SAIS} Review of International Affairs}, 35(1):73--86, 2015.

\bibitem{sullivan_2017}
J.~E. Sullivan and D.~Kamensky.
\newblock How cyber-attacks in ukraine show the vulnerability of the u.s. power
  grid.
\newblock {\em The Electricity Journal}, 30(3):30 -- 35, 2017.

\bibitem{mishra_2020}
S.~Mishra, K.~Anderson, B.~Miller, K.~Boyer, and A.~Warren.
\newblock Microgrid resilience: A holistic approach for assessing threats,
  identifying vulnerabilities, and designing corresponding mitigation
  strategies.
\newblock {\em Applied Energy}, 264:114726, 2020.

\bibitem{ji_2017}
C.~{Ji}, Y.~{Wei}, and H.~V. {Poor}.
\newblock Resilience of energy infrastructure and services: Modeling, data
  analytics, and metrics.
\newblock {\em Proceedings of the {IEEE}}, 105(7):1354--1366, Jul 2017.

\bibitem{chen2014short}
Kuilin Chen and Jie Yu.
\newblock Short-term wind speed prediction using an unscented kalman filter
  based state-space support vector regression approach.
\newblock {\em Applied Energy}, 113:690--705, 2014.

\bibitem{mukherjee_2018}
S.~Mukherjee, R.~Nateghi, and M.~Hastak.
\newblock A multi-hazard approach to assess severe weather-induced major power
  outage risks in the {U.S.}
\newblock {\em Reliability Engineering \& System Safety}, 175:283--305, 2018.

\bibitem{adderly_2016}
S.~Adderly.
\newblock {\em Reviewing Power Outage Trends, Electric Reliability Indices and
  Smart Grid Funding}.
\newblock PhD thesis, University of Vermont, 2016.

\bibitem{larsen_2020}
P.~H. Larsen, M.~Megan~Lawson, K.~Hamachi-LaCommare, and J.~H. Eto.
\newblock Severe weather, utility spending, and the long-term reliability of
  the {U.S.} power system.
\newblock {\em Energy}, 198:117387, 2020.

\bibitem{oe417form}
{OE-417} electric emergency incident and disturbance report.
\newblock {https://www.oe.netl.doe.gov/OE417/Form/Home.aspx}.
\newblock Accessed: 2020-10-13.

\bibitem{oe417data}
Electric disturbance events ({OE}-417).
\newblock {https://www.oe.netl.doe.gov/oe417.aspx}.
\newblock Accessed: 2020-10-13.

\bibitem{oe417_annualsummary}
Electric disturbance events ({OE}-417) annual summaries.
\newblock {https://www.oe.netl.doe.gov/OE417\_annual\_summary.aspx}.
\newblock Accessed: 2020-10-13.

\bibitem{eia}
Annual electric power industry report, form {EIA}-861 detailed data files.
\newblock {https://www.eia.gov/electricity/data/eia861/}.
\newblock Accessed: 2020-10-13.

\bibitem{qer_017}
Quadrennial Energy Review~Task Force.
\newblock Transforming the nation’s electricity system: The second
  installment of the {QER}.
\newblock {\em Quadrennial Energy Review Task Force}, 2017.

\bibitem{tibshirani1996regression}
R.~Tibshirani.
\newblock Regression shrinkage and selection via the lasso.
\newblock {\em Journal of the Royal Statistical Society: Series B
  (Methodological)}, 58(1):267--288, 1996.

\bibitem{data_2018}
S.~Mukherjee, R.~Nateghi, and M.~Hastak.
\newblock Data on major power outage events in the continental {U.S.}
\newblock {\em Data in Brief}, 19:2079 -- 2083, 2018.

\bibitem{larsen_cost_2018}
Peter~H. Larsen, Brent Boehlert, Joseph Eto, Kristina Hamachi-LaCommare, Jeremy
  Martinich, and Lisa Rennels.
\newblock Projecting future costs to u.s. electric utility customers from power
  interruptions.
\newblock {\em Energy}, 147:1256 -- 1277, 2018.

\bibitem{rocchetta_2018}
Roberto Rocchetta, Enrico Zio, and Edoardo Patelli.
\newblock A power-flow emulator approach for resilience assessment of
  repairable power grids subject to weather-induced failures and data
  deficiency.
\newblock {\em Applied Energy}, 210:339--350, 2018.

\bibitem{us_regions}
Geography~Division of~the U.S. Census~Bureau.
\newblock Census regions and divisions of the united states.
\newblock
  {https://www2.census.gov/geo/pdfs/maps\-data/maps/reference/us\_regdiv.pdf}.
\newblock Accessed: 2020-10-13.

\bibitem{wh_energy_2013}
The~White House, President’s~Council of~Economic~Advisers, U.S.~Department
  of~Energy’s Office~of Electricity~Delivery, Energy Reliability, White
  House~Office of~Science, and Technology.
\newblock Economic benefits of increasing electric grid resilience to weather
  outages.
\newblock {https://www.energy.gov/sites/prod/files/2013/08/f2/Grid Resiliency
  Report\_FINAL.pdf}.
\newblock Accessed: 2020-10-13.

\bibitem{wang_2018}
N.~Wang.
\newblock Transactive control for connected homes and neighbourhoods.
\newblock {\em Nature Energy}, 3(11):907--909, September 2018.

\bibitem{bureau_2019}
United States~Census Bureau.
\newblock {U.S.} census bureau quickfacts: Seattle city, {Washington}; {United
  States}.
\newblock
  {https://www.census.gov/quickfacts/fact/table/seattlecitywashington,US/PST045219}.
\newblock Accessed: 2020-10-13.

\bibitem{scl}
Seattle~City Light.
\newblock What is advanced metering?
\newblock {https://www.seattle.gov/light/ami/metering.asp}.
\newblock Accessed: 2020-10-13.

\bibitem{kroman_2017}
D.~Kroman.
\newblock Facing cost overruns, city light quietly pares smart-meter project.
\newblock {\em Crosscut}, May 2017.

\bibitem{mazur_2019}
Christoph Mazur, Yannick Hoegerle, Maria Brucoli, Koen {van Dam}, Miao Guo,
  Christos~N. Markides, and Nilay Shah.
\newblock A holistic resilience framework development for rural power systems
  in emerging economies.
\newblock {\em Applied Energy}, 235:219--232, 2019.

\bibitem{inklaar_2018}
R.~Inklaar, H.~de~Jong, J.~Bolt, and J.~van Zanden.
\newblock Rebasing `maddison': new income comparisons and the shape of long-run
  economic development.
\newblock Technical report, Groningen Growth and Development Centre, University
  of Groningen, 2018.

\bibitem{scl_undergrounding}
Seattle~City Light.
\newblock Neighborhood underground conversion program.
\newblock https://www.seattle.gov/light/underground.asp.
\newblock Accessed: 2020-10-13.

\bibitem{jufri_2019}
Fauzan~Hanif Jufri, Victor Widiputra, and Jaesung Jung.
\newblock State-of-the-art review on power grid resilience to extreme weather
  events: Definitions, frameworks, quantitative assessment methodologies, and
  enhancement strategies.
\newblock {\em Applied Energy}, 239:1049--1065, 2019.

\bibitem{graham_2010}
S.~Graham.
\newblock {\em Disrupted cities: when infrastructure fails}.
\newblock Routledge, New York, 2010.

\bibitem{li_2010}
H.~{Li}, L.~A. {Treinish}, and J.~R.~M. {Hosking}.
\newblock A statistical model for risk management of electric outage forecasts.
\newblock {\em IBM Journal of Research and Development}, 54(3):8:1--8:11, May
  2010.

\bibitem{schmidthaler_2016}
M.~Schmidthaler and J.~Reichl.
\newblock Assessing the socio-economic effects of power outages ad hoc.
\newblock {\em Computer Science - Research and Development}, 31(3):157--161,
  March 2016.

\bibitem{1366}
{IEEE} guide for electric power distribution reliability indices.
\newblock {\em IEEE Std 1366-2012 (Revision of IEEE Std 1366-2003)}, pages
  1--43, 2012.

\bibitem{hann_2011}
N.~{Hann} and C.~{Daly}.
\newblock Investigation of the 2.5 beta methodology.
\newblock {\em IEEE Transactions on Power Systems}, 26(4):2577--2578, Nov 2011.

\end{thebibliography}


\begin{thebibliography}{1}

\bibitem{oe417_reporting}
Office of~Electricity~Delivery and Energy Reliability.
\newblock {OE}-417 electric emergency incident and disturbance report.
\newblock {https://www.frcc.com/disturbance/Shared Documents/Disturbance
  Reporting/DOE OE\-417 Instructions.pdf}.
\newblock Accessed: 2020-10-13.

\bibitem{lacommare_2018}
K.~Hamachi-LaCommare, J.~H. Eto, L.~N. Dunn, and M.~D. Sohn.
\newblock Improving the estimated cost of sustained power interruptions to
  electricity customers.
\newblock {\em Energy}, 153:1038 -- 1047, 2018.

\bibitem{ellis_2008}
S.~Ellis.
\newblock The case for profitable proximity.
\newblock
  {https://www.supplychainquarterly.com/articles/189-the-case-for-profitable-proximity}.
\newblock Accessed: 2020-10-13.

\bibitem{peterson_2017}
D.~Peterson.
\newblock Link between growth in economic activity and electricity use is
  changing around the world.
\newblock {https://www.eia.gov/todayinenergy/detail.php?id=33812}.
\newblock Accessed: 2020-10-13.

\end{thebibliography}
\end{document}

\begin{frontmatter}
\title{Supplementary Material for U.S. Power Resilience for 2002--2019}
\end{frontmatter}

% \author[mymainaddress]{Aman Ankit}

% \author[mymainaddress]{Zhanlin Liu}

% \author[scottaddress]{Scott B. Miles}

% \author[mymainaddress]{Youngjun Choe\corref{mycorrespondingauthor}}
% \cortext[mycorrespondingauthor]{Corresponding author}
% \ead{ychoe@uw.edu}

% \address[mymainaddress]{Department of Industrial and Systems Engineering, University of Washington, Seattle, WA 98105, USA}
% \address[scottaddress]{Department of Human Centered Design \& Engineering, University of Washington, Seattle, WA 98105, USA}
%\end{frontmatter}
% Supplementary notes

\section{Supplementary Notes}
\subsection{Supplementary Note 1: Federal reporting mandates for filing OE-417}
\label{supp:sec1}
The document that outlines instructions on filling the OE-417 form states that electric utilities must report every outage without exclusions within 1 hour, 6 hours, or 48 hours depending on the severity due to Section 13(b) of the Federal Energy Administration Act of 1974  \cite{oe417_reporting}. Failure to comply will result in sanctions and fines every day that it is not reported. It is also a criminal offense to “knowingly and willingly to make to any Agency or Department of the United States any false, fictitious, or fraudulent statements as to any matter within its jurisdiction” \cite{oe417_reporting}. There are multiple entities that report the outage although there are some exclusions based on the entity. Electric utilities as mentioned earlier may not exclude anything, but balancing authorities can be exempt from reporting the count of customers unless it is for a Final Report \cite{oe417_reporting}. %There are too many risks to not file a report and 
It is too risky for an entity to not file a report because multiple entities file the outage report that can be cross-referenced to ensure the most accurate published data. Since we believe we have the most accurate data available due to the regulations, we conclude that gray states are actually the most resilient in the category of cause.

\subsection{Supplementary Note 2: Analysis on SAIFI reliability metric for outages caused by natural events}
\label{supp:sec2}
The LASSO model for SAIFI natural had the same significant predictors as SAIFI state: population percentage in urban areas (regression coefficient: \num{-4.5031e-2}, standard error: \num{1.44169e-2}, $p$-value: 0.03123) and population density of rural areas (regression coefficient: \num{1.90448e-2}, standard error: \num{7.4363e-2}, $p$-value: 0.01386). The logic used to construct the undergrounding hypothesis for SAIFI state can also be applied here with an added explanation to relate it to natural causes as it is known that one of the main reasons electrical companies are called to repair power lines is due to falling tree branches. Undergrounding, which is less common in rural areas and more common in urban areas, eliminates this reason, thus reducing the SAIFI values.

\subsection{Supplementary Note 3: Analysis on CAIDI reliability metric for all types of outages}
\label{supp:sec3}
CAIDI state's most significant predictor was the percentage of power consumption by industrial customers (regression coefficient: \num{-0.06817}, standard error: \num{0.03973}, $p$-value: \num{0.09309}). A large percentage of industrial customers can indicate that the electrical utility companies may receive a proportionally higher revenue from them as they can be inferred to consume more electricity. Areas that have a high percentage of industrial customers may signify higher demand with lower tolerance for downtime because the cost of a power outage in the industrial sector versus residential or commercial sectors is much higher  \cite{lacommare_2018}. As a result, utility companies may invest more in preparedness and resources to combat natural outage types and ensure that the outage duration is limited. Furthermore, a large percentage of industrial customers can mean that the region is more attractive to other companies due to advantages regarding “profitable proximity” in relation to the supply chain thus creating a better economic region and providing more revenue for the utility companies that they can invest in preparedness and response time \cite{ellis_2008}. If the theory of profitable promixity maintains true in these regions, utility companies may have built their infrastructure to prevent outages from affecting a large population as they expected the economies of their regions grow \cite{peterson_2017} and attract residents as well.

\subsection{Supplementary Note 4: Analysis on CAIDI reliability metric for outages caused by operational maintenance factors}
\label{supp:sec4}
CAIDI operations had population density of urban areas (regression coefficient: \num{4.372e-4}, standard error: \num{1.360e-4}, $p$-value: 0.00236) as its only significant predictor with a positive correlation. One hypothesis to explain this relationship is based on the increased complexity to maintain the power grid as the urban area's population density increases. For example, a utility company would need to coordinate with a larger number of entities for scheduling an outage, and thus the utility would want to perform multiple tasks during each outage, prolonging the outage duration per each event (i.e. increased CAIDI). Another hypothesis may be due to undergrounding. Previously in the discussion for SAIFI operational, we hypothesized that as regions begin urbanizing, there may be an increased demand in undergrounding. A region that is rapidly growing in population and urbanizing, as indicated by an increasing population density, may have increased outage durations for two potential reasons. As the sudden demand for undergrounding grows then utility companies may not have the proper experience and failures to learn from prior events. Procedures may not have been to the same standard as those electric companies that have high experience in maintaining underground infrastructure. The operational maintenance crew may face undocumented events that were a result of a lack of clear and complete procedures due to inexperience of the utility company resulting in increased outage duration. A second reason may relate to the lack of experienced new hires. As the demand for rapid expansion and infrastructure grows, utility companies may see an uptick in new hires who will be slower at operational maintenance, thus increasing CAIDI.

% Supplementary figures
\section{Supplementary Figures and Tables}

\begin{figure}[h!]
    \centering
    \includegraphics[width=0.95\textwidth]{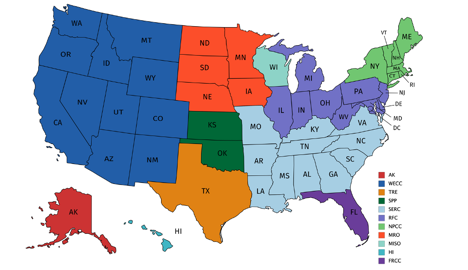}
    \caption{NERC regions in the U.S.}
    \label{supp_fig:1}
\end{figure}

\begin{figure}[h!]
    \centering
    \includegraphics[width=0.95\textwidth]{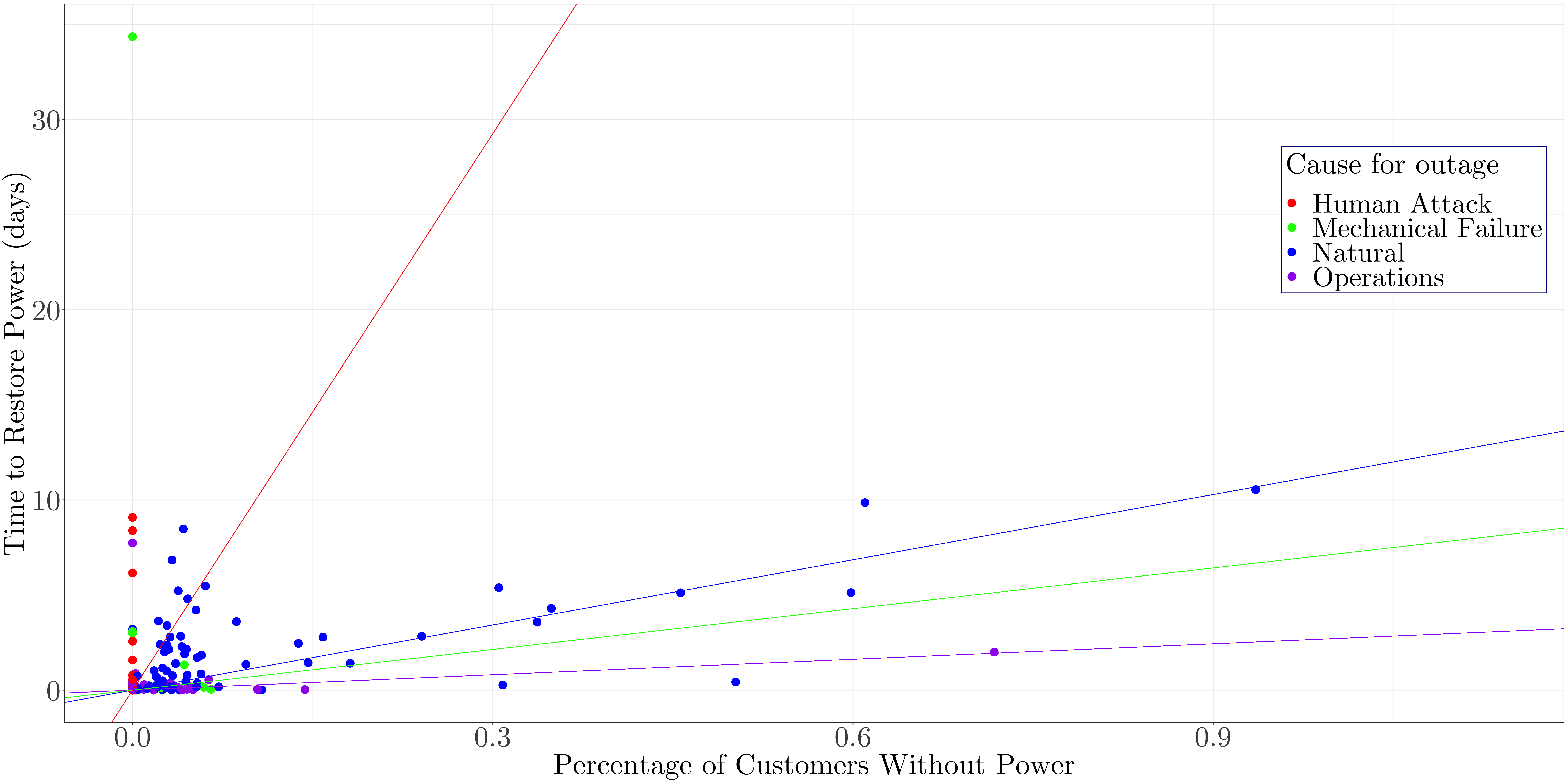}
    \caption{Power outage events with a regression line for each outage cause for the state of California for 2002--2019.}
    \label{supp_fig:2}
\end{figure}

% Supplementary tables
%\section{Supplementary Tables}

\begin{table}[h!]
\centering
\begin{threeparttable}
\caption{Measure of influence using various methods for California for 2002--2019}
\label{tab:influence}
\begin{tabular}{SSSSSS[table-format=3.2]}
\toprule
\textbf{Observation} & \textbf{\thead{Difference\\of Betas}} & \textbf{\thead{Difference\\of Fit}} & \textbf{\thead{Coefficient of\\Variation Ratio}} & \textbf{Cook's Distance} & \textbf{\thead{Hat Diagonal\\Values}} \\ \midrule
\text{November}, \num{2002}  & 0.00 & 0.00   & 1.05 * & 0.00   & 0.04 *   \\
\text{December}, \num{2005} & -0.23 & -0.23   & 1.14 * & 0.05   & 0.12 * \\
\text{December}, \num{2006} & 0.24 & 0.24   & 1.02 & 0.06   & 0.03 * \\
\text{July}, \num{2006} & 0.08 & 0.08   & 1.08 * & 0.01   & 0.07 * \\
\text{February}, \num{2007} & 0.03 & 0.03   & 1.03 * & 0.00   & 0.02 \\
\text{January}, \num{2008} & 0.52 & 0.52  *   & 1.36 * & 0.27   & 0.27 * \\
\text{October}, \num{2009} & -0.28 & -0.28 *  & 1.02 & 0.08   & 0.03 * \\
\text{January}, \num{2010} & 0.85  & 0.85 *  & 1.08 * & 0.70 *   & 0.11 * \\
\text{January}, \num{2010} & 0.06  & 0.06  & 0.97 * & 0.00   & 0.00 \\
\text{September}, \num{2011} & -1.42 *  & -1.42 *  & 1.09 * & 1.85 *  & 0.15 * \\
\text{March}, \num{2011} & 0.06  & 0.06  & 0.96 * & 0.00   & 0.00   \\
\text{December}, \num{2011} & 0.07  & 0.07    & 0.89 * & 0.00   & 0.00   \\
\text{May}, \num{2014} & -0.74 & -0.74 * & 1.02   & 0.53 * & 0.07 * \\
\text{January}, \num{2017} & 0.05  & 0.05    & 0.94 * & 0.00   & 0.00   \\
\text{February}, \num{2017} & 0.09  & 0.09    & 0.94 * & 0.01   & 0.00   \\
\text{October}, \num{2017} & 0.10  & 0.10    & 0.82 * & 0.01   & 0.00   \\
\bottomrule
\end{tabular}
  \begin{tablenotes}
      \small
      \item \textit{Note}: For each measure of influence, observations with a starred number were deemed influential.
    \end{tablenotes}
  \end{threeparttable}
\end{table}

\begin{table}[h!]
\caption{Information extracted from Form OE-417.}
\label{Supple table 2}
\begin{tabular}{lr}
\toprule
\textbf{Relevant columns from OE-417} & \textbf{Definition}                             \\ \hline
Date Event Began                      & The day/month/year of the outage                \\
Time Event Began                      & The time (am/pm) of the outage                  \\
Date of Restoration                   & The day/month/year when the outage was resolved \\
Time of Restoration                   & The time (am/pm) when the outage was resolved   \\
Area Affected                         & The state (sometimes also lists county)         \\
Event Type                            & The reason for the outage                       \\
Number of Customers Affected          & The number of customers that lost power         \\ \hline
\end{tabular}
\end{table}

\begin{table}[h!]
\caption{Information extracted from Form EIA-861.}
\label{Supple table 3}
\begin{adjustbox}{width=0.95\textwidth}
\begin{tabular}{cc}
\toprule
\textbf{Relevant columns from EIA-861} & \textbf{Definition}                                                              \\ \midrule
State                                  & The state that the utility company belongs to                                    \\ 
\multirow{ 2}{*}{Number of Customers} &  The number of customers\\
& that purchased electricity in the specified utility name
% Numbers of Customers                   & The number of customers that purchased electricity in the specified utility name
 \\ \bottomrule
\end{tabular}
\end{adjustbox}
\end{table}

Supplementary Tables 4, 5, and 6 are contained in ``Summary of Selected Significant Predictors for CAIDI.html", ``Summary of Selected Significant Predictors for SAIDI.html", and ``Summary of Selected Predictors for SAIFI.html", respectively.

\clearpage
\bibliographystyle{unsrt}
\bibliography{references}